\pdfoutput=1

\documentclass[reprint,amsmath,amssymb,floatfix,aps,longbibliography]{revtex4-1}
\usepackage{graphics,graphicx,float}

\usepackage[pdftex,
            pdfauthor={Steven A. Frank},
            pdftitle={},
            pdfsubject={Natural selection. III. Selection versus transmission and the levels of selection}]{hyperref} 

\def\citeyear{\citep}
\def\autocite{\citep}

\newcommand{\zbar}{\bar{z}}
\newcommand{\wbar}{\bar{w}}
\newcommand{\cov}{{\hbox{\rm Cov}}}

\newcommand{\R}{{\mathcal{R}}}

\newcommand{\Ga}{\alpha}
\newcommand{\Gb}{\beta}
\newcommand{\Gd}{\delta}
\newcommand{\GD}{\Delta}
\newcommand{\Ge}{\epsilon}
\newcommand{\Gg}{\gamma}

\newcommand{\Gm}{\mu}

\newcommand{\Gt}{\tau}

\DeclareMathOperator{\E}{E}

\newcommand{\Eq}[1]{Eq.~(\ref{eq:#1})}

\newcommand{\Fig}[1]{Fig.~\ref{fig:#1}}

\newcommand{\boldrule}{\hrule height 1.2pt}
\newcommand{\noterule}{\medskip\boldrule\medskip}	

\newcount\BoxNum \BoxNum 1\relax
\makeatletter
\newcommand{\boxlabel}[1]{%
  \protected@write \@auxout {}{\string \newlabel {box:#1}{{\the\BoxNum}{\thepage}{\noexpand\relax}%
  	{\@ifundefined{hyper@@anchor}{\relax}{box.\the\BoxNum}}%
  	{}}}%
  \@ifundefined{hyper@@anchor}{}{\hypertarget{box.\the\BoxNum}{}}%
  \advance\BoxNum 1\relax}
\makeatother
\newcommand{\Boxx}[1]{Box~\ref{box:#1}}
\newcommand{\BoxLabel}{Box~\the\BoxNum}

\begin{document}

\title{Natural selection. III. Selection versus transmission and the levels of selection}

\author{Steven A.\ Frank}
\email[email: ]{safrank@uci.edu}
\homepage[homepage: ]{http://stevefrank.org}
\affiliation{Department of Ecology and Evolutionary Biology, University of California, Irvine, CA 92697--2525  USA}

\begin{abstract}

George Williams defined an evolutionary unit as hereditary information for which the selection bias between competing units dominates the informational decay caused by imperfect transmission.  In this article, I extend Williams' approach to show that the ratio of selection bias to transmission bias provides a unifying framework for diverse biological problems.  Specific examples include Haldane and Lande's mutation-selection balance, Eigen's error threshold and quasispecies, Van Valen's clade selection, Price's multilevel formulation of group selection, Szathm{\'a}ry and Demeter's evolutionary origin of primitive cells, Levin and Bull's short-sighted evolution of HIV virulence, Frank's timescale analysis of microbial metabolism, and Maynard Smith and Szathm{\'a}ry's major transitions in evolution. The insights from these diverse applications lead to a deeper understanding of kin selection, group selection, multilevel evolutionary analysis, and the philosophical problems of evolutionary units and individuality\footnote{\href{http://dx.doi.org/10.1111/j.1420-9101.2011.02431.x}{doi:\ 10.1111/j.1420-9101.2011.02431.x} in \textit{J. Evol. Biol.}}\footnote{Part of the Topics in Natural Selection series. See \Boxx{preface}.}.

\end{abstract}

\maketitle

\begin{quote}
In evolutionary theory, a gene could be defined as any hereditary information for which there is a $\ldots$ selection bias equal to several or many times its rate of endogenous change \autocite[p.~44]{williams66adaptation}.
\end{quote}

\section*{Introduction}

Natural selection increases inherited information about environmental challenge.  Against selection, imperfect transmission reduces inherited information.  Many problems in biology come down to understanding the relative balance between selection and imperfect transmission.  

A clear understanding of selection and transmission requires greater precision with regard to abstract notions such as \textit{inherited information}. However, before heading off in pursuit of abstract theory, it pays to have some simple examples in mind.  Those simple examples define the challenges for deeper theory.

In this paper, I work through several examples that turn on the relative strength of selection and imperfect transmission: Haldane \citeyear{haldane27a-mathematical} and Lande's \citeyear{lande75the-maintenance} balance between selection and mutation, Eigen's \citeyear{eigen92steps} error threshold and quasispecies, Van Valen's \citeyear{van-valen75group} multilevel analysis of clade selection, Price's \citeyear{price72extension} multilevel analysis of group selection, Szathm{\'a}ry and Demeter's \citeyear{szathmary87group} stochastic corrector model of early cellular evolution, Levin and Bull's \citeyear{levin94short-sighted} short-sighted model of parasite evolution, Frank's \citeyear{frank10the-trade-off} timescale model of microbial metabolism, and Maynard Smith and Szathm{\'a}ry's \citeyear{maynard-smith95the-major} major transitions in evolution.  

Others have pointed out similarities between some of these examples \autocite{maynard-smith95the-major,michod06cooperation,okasha06evolution}.  However, the broad unity with regard to selection and transmission is sometimes lost.  In addition, the key role of timescale, although often noted, has not always been linked to selection and transmission in a simple and general way.

Williams' \citeyear{williams66adaptation} quote emphasizes timescale: the opposition between \textit{selection bias} and \textit{rate of endogenous change.}  An entity can be shaped by natural selection only to the extent that the informational gain by natural selection is not overwhelmed by the relative rate of informational decay by imperfect transmission.  The balance between selection and decay often turns on the relative timescales over which those forces operate.

\section*{The decay of transmission fidelity}

Many processes reduce the similarity between ancestor and descendant.  In classical genetics, mutation changes the intrinsic quality of an allele during transmission.  Mixing of alleles reduces transmission fidelity because of interactions with the changed combination of other alleles.  Internal selection changes the frequency of alleles within individuals, altering the similarity between ancestor and descendant.  

Internal selection may occur within a pool of allele copies, in which certain alleles express traits that cause their frequency to increase against their neighbors \autocite{burt08genes}.  For example, shortened mitochondrial genomes in certain yeast replicate faster than full genomes.  The shortened genomes can rise in frequency within cells, even though they reduce individual-level fitness.  In diploid Mendelian genetics, internal selection arises when traits increase allelic transmission to offspring to greater than the standard Mendelian probability of one-half.

Mutation or mixing of alleles may, in some cases, cause unbiased change during transmission.  Unbiased change decays transmission fidelity, but does not affect the direction of evolution for the average value of traits.  Unbiased change can increase the variation in traits by causing random fluctuations in the characters expressed by descendants.  Under stabilizing selection, the amount of 

\begin{figure}[H]
\begin{minipage}{\hsize}
\parindent=15pt
\noterule
{\bf \noindent\BoxLabel. Topics in the theory of natural selection}
\noterule
This article is part of a series on natural selection.  Although the theory of natural selection is simple, it remains endlessly contentious and difficult to apply.  My goal is to make more accessible the concepts that are so important, yet either mostly unknown or widely misunderstood.  I write in a nontechnical style, showing the key equations and results rather than providing full derivations or discussions of mathematical problems.  Boxes list technical issues and brief summaries of the literature.  
\noterule
\end{minipage}
\end{figure}
\boxlabel{preface}

\noindent variation may be shaped by a balance between an increase caused by fluctuations in transmission and a decrease caused by selection removing fluctuations from the favored value \autocite{lande75the-maintenance}.

Biased mutation or internal selection causes a directional change during transmission.  When the directional change during transmission opposes selection between individuals or groups, the balance between selection and transmission influences the average value of traits.

\section*{Selection versus transmission}

Total evolutionary change can be partitioned into components of selection and transmission:
\begin{center}
  Total change = $\GD$selection $+$ $\GD$transmission,
\end{center}
in which the symbol $\GD$ means \textit{the change caused by the process of} or \textit{the change in the quantity of} depending on context.  This partition of total change into selection and transmission is so important that it is worthwhile to express the partition with symbols.  The symbolic form allows us to look at variations of the partition and the consequences for understanding evolutionary process (\Boxx{price}). 

\textit{Total change} can be expressed by the change in the average value of some trait.  Let $\GD\zbar$ be the change in the average trait value.  Do not be misled by the word \textit{average}.  We can consider the average of the squared deviations of a trait to measure the variance, or the average of the product of different characters to measure correlations, or the average frequency of an allele in the population, or any other expression leading to some quantity:  $\GD\zbar$ is the change in whatever quantity we choose.  We write total evolutionary change as $\wbar\GD\zbar$, where $\wbar$ is average fitness.  Average fitness accounts for the total numbers of births and deaths, allowing us to express selection and transmission directly in proportion to total change (see \Boxx{price}).

Express the change caused by selection as $\GD S$ and the change caused by transmission as $\GD\Gt$. Then the total change in symbols is
\begin{equation}\label{eq:total}
  \wbar\GD\zbar=\GD S + \GD\Gt.
\end{equation}

\begin{figure}[H]
\begin{minipage}{\hsize}
\parindent=15pt
\noterule
{\bf \noindent\BoxLabel. Price's selection and transmission}
\noterule
The Price equation provides a useful separation between selection and transmission \autocite{price70selection,price72extension,hamilton75innate}. Much literature and misunderstanding descend from the Price equation.  I will treat the topic fully in a later article. Here, I briefly summarize the essential concepts. My previous publications related to the Price equation provide further background \autocite{frank95george,frank97the-price,frank98foundations}.  Other key references lead into the broader literature \autocite{wade85soft,heisler87a-method,michod97cooperation,grafen02a-first,rice04evolutionary,okasha06evolution,gardner08the-price}.

I used the Price equation as the basis for \Eq{total} in the text. The Price equation may be written as
\begin{equation*}
  \wbar\GD\zbar = \cov(w,z) + \E(w\GD z).
\end{equation*}
Comparing to \Eq{total}, the selection bias is $\GD S = \cov(w,z)$.  This simply says that the selection bias is the association between fitness and character value, where association is expressed by the covariance.  The transmission bias is $\GD\Gt = \E(w\GD z)$.  This says that the transmission bias is the average (expectation) of the change in character value, $\GD z$, between parent and offspring.  The individual parent-offspring biases in transmission are weighted by parental fitness, $w$.  If, for example, a parent reproduces little, then that parent's transmission bias contributes little to the average transmission bias in the population. 

The expression for selection in \Eq{sV1} is derived as $\GD S = \cov(w,z) = \Gb_{wz} V_z$, because the covariance of $w$ and $z$ is the product of the regression coefficient, $\Gb$, of $w$ on $z$, and the variance of $z$.  Define $s_z=\left|\Gb_{wz}\right|$, and apply a minus sign when $\Gb_{wz}<0$ to obtain \Eq{sV2}.  See \textcite{frank97the-price} for the interpretation of these terms in the Price equation.

In the mutation-selection balance models, either $z\equiv q$ is allele frequency or $z$ is the squared deviation of a trait from the optimum. In either case, $z$ is always positive, and the association between fitness and character value is negative. Thus, $-s_z=\Gb_{wz}$, and we can express fitness in terms of the regression form 
\begin{equation}\label{eq:fitRegress}
  \E(w|z) = 1+\Gb_{wz}z = 1-s_z z.  
\end{equation}
Here, I set maximum fitness to one.  Any proportional change in maximum fitness is matched by the same proportional change in the regression coefficient, so the expression can be scaled arbitrarily. From this regression expression, the average of $s_z z$ must be less than one, otherwise average fitness drops below zero and mutational decay dominates selection, causing loss of heritable information or ``mutational meltdown'' \autocite{lynch93the-mutational}.

Note that the regression expression $\E(w|z) = 1-\Gb_{wz}z$ does not require a linear relation between character value and fitness.  Rather, $\Gb_{wz}$ is simply the best least squares fit of fitness to trait value given the actual pattern by which trait values associate with fitness.  
\noterule
\end{minipage}
\end{figure}
\boxlabel{price}

\noindent Any evolutionary problem can be expressed in this way.  But whether it is useful to do so depends on the particular problem and, to some extent, on one's preference between alternative ways to partition total change into various components.

Selective improvement often pushes traits in the opposite direction from transmission decay.  The balance between these opposing forces occurs when the total change is zero
\begin{equation}\label{eq:total0}
  \wbar\GD\zbar=\GD S + \GD\Gt=0,
\end{equation}
which also means that at an equilibrium balance 
\begin{equation}\label{eq:totalBalance}
  \GD S = -\GD\Gt.
\end{equation}
This equation provides the ultimate expression of a balance between selective improvement and transmission decay \autocite{frank90the-distribution}. 

We can often write the change caused by selection as 
\begin{equation}\label{eq:sV1}
  \GD S = s_zV_z,
\end{equation}
where $s_z$ is the selective intensity on the character $z$, and $V_z$ is the variance in the character $z$ under selection (see \Boxx{price}).  If selection causes a decrease in the character, we would instead write 
\begin{equation}\label{eq:sV2}
  \GD S = -s_zV_z
\end{equation}
to express the negative contribution of selection to the change in character. Using these expressions for the change caused by selection in \Eq{totalBalance}, we obtain the equilibrium variance under a balance between selection and transmission as
\begin{equation}\label{eq:varianceBalance}
	V_z = \left|\frac{\GD\Gt}{s_z}\right|.
\end{equation}
The absolute value is used because $s_z$ and $V_z$ are always positive, whereas $\GD\Gt$ may be positive or negative depending on whether the transmission bias increases or decreases the trait.  The key point is that when the opposing forces of selection and transmission are in balance, we have this simple expression for the variance of a character. 

\section*{A measure of selection versus transmission}

How exactly should we interpret Williams' phrase ``hereditary information for which there is a $\ldots$ selection bias equal to several or many times its rate of endogenous change''?  We could evaluate the strength of selection bias relative to transmission bias to obtain a simple measure for the ratio, $\R$, between the forces. In particular, when the two terms oppose each other, we may write
\begin{equation}\label{eq:dominance}
	\R = \log\left(-\frac{\GD S}{\GD\Gt}\right).
\end{equation}
The negative sign appears because the opposing directions of change for $\GD S$ and $\GD\Gt$ mean that these terms have opposite signs.  The negative sign makes the ratio positive.  The logarithmic scaling imposes symmetry about zero.  The ratio is zero when the two forces are \hbox{\hskip-\fill}\break

\begin{figure}[H]
\begin{minipage}{\hsize}
\parindent=15pt
\noterule
{\bf \noindent\BoxLabel. What are groups?}
\noterule
One must distinguish between two aspects. On the one hand, the fundamental theory works perfectly for essentially any conception of groups of alleles, individuals, or other entities.  The groups do not require clear delineation, temporal continuity, or biologically meaningful interaction. \textit{Selection within groups} simply means the differential success between entities in the group, however that differential success arises. \textit{Transmission bias} simply means the fitness weighted change in character value between the entities in the group and their descendants.  No restriction is placed on how the descendants themselves are arranged into groups.

On the other hand, most potential groupings have no biological meaning.  One naturally prefers groups defined by direct interaction, temporal continuity, shared interest, and so on.  Much literature debates alternative conceptions of meaningful groups \autocite{maynard-smith76group,wilson89reviving,michod97evolution,gardner09capturing}.  Difficulty occurs because the relative value of alternative views varies with biological context, intellectual goal, and subjective bias about what is ultimately meaningful.  Such undecidable alternatives attract endless debate and commentary.  

Discussion of biologically meaningful alternatives can lead to improved understanding as the weight of evidence accumulates for certain views.  However, that discussion has often sought absolute conclusions, when in fact context and subjective aspects necessarily play a role.  In my view, one needs to keep in mind both the fundamental truth of the universal theory and the nuance of changing context and meaning in application.  With both perspectives in mind, one never loses way.
\noterule
\end{minipage}
\end{figure}
\boxlabel{groups}

\noindent equal, as in the balance condition of \Eq{totalBalance}.  Increasingly positive values arise from greater dominance of selection bias, whereas increasingly negative values arise from greater dominance of transmission bias. Later examples illustrate the application of this ratio.

\section*{Multilevel selection}

The individual typically comprises a group of alleles. In some cases, selection may occur between alleles within the individual.  That selection within individuals creates a transmission bias between ancestors and descendants, because the sample of alleles transmitted to descendants is changed by selection between alleles within the ancestor.  The total change can be expressed as selection between individuals plus the transmission bias created by selection within individuals.  In this case, we can think of selection and transmission as the combination of two levels of selection \autocite{price72extension,hamilton75innate}.  

Now consider a population of individuals structured into groups. The total change may be partitioned into selection between groups and the transmission bias between an ancestral group and the descendants derived from that group (\Boxx{groups}).  Selection between individuals within the group will often strongly influence transmission bias, because selection within the group changes the composition of traits that are transmitted to descendants of that group.  The total change can be expressed primarily as selection between groups and the transmission bias created by selection within groups.  Once again, we can think of selection and transmission as the combination of two levels of selection.  

\section*{Timescale}

The balance between selection and transmission depends on the rate of selection between groups relative to the rate of endogenous change within groups.  Timescale influences the relative rates.  

Consider, for example, an increasing number of rounds of selection within groups for each round of selection between groups.  If there is some limit to the ultimate size of groups, then the transmission bias caused by selection within groups increasingly dominates the selection between groups \autocite{wilson81the-evolution}.  Similarly, an increase in the number of rounds of replication within a lineage relative to the timescale of selection between lineages causes relatively greater mutation and decay during transmission compared with the selection bias. For example, the male mutation rate appears to be greater than the female mutation rate in several animal species, probably because of the greater number of replications per generation in the male germline \autocite{nachman00estimate}.  

It seems obvious that a relatively greater time for selection bias or transmission decay enhances the relative strength of a process.  However, the simplicity of partitioning total change into selection and transmission in relation to timescale is not always developed clearly.  By going through the examples properly, we can recover the simple conceptual unity that helps to explain a wide variety of biological problems.

\section*{Balance between selection and mutation}

Perhaps the most basic of all evolutionary theory concerns the balance between selection and mutation \autocite{haldane27a-mathematical}.  From \Eq{total}, let the trait $\zbar\equiv q$ be the frequency of a deleterious allele.  The equilibrium balance between selection and mutation occurs when the rate at which selection removes deleterious alleles equals the rate at which mutation adds new deleterious alleles. From \Eq{totalBalance}, the balance occurs when $\GD S=-\GD\Gt$.  From \Eq{varianceBalance}, we can also express that balance as
\begin{equation}\label{eq:mutVar}
	V_q = \frac{\GD\Gt}{s_q},
\end{equation}
where $V_q$ is the variance in allele frequency, and $s_q$ is the selective intensity on allele frequency.  In this case, mutation increases the frequency of the mutant allele, so $\GD\Gt$ is positive and we do not need to use absolute values.  

\subsection*{Classic results of population genetics}

Suppose mutation changes a normal allele into a deleterious allele.  Once an allele has become deleterious, it cannot mutate back into a normal allele.  Let the mutation rate of normal alleles be $\Gm$. Normal alleles occur at frequency $1-q$. Thus, the change in the number of mutant alleles caused by transmission bias is in proportion to $\GD\Gt=\Gm(1-q)$.  

Selection reduces the reproductive success of mutant alleles by the selective intensity, $s_q\equiv s$. The variance in allele frequency is $V_q=q(1-q)$, the standard binomial expression for variance when sampling a single allele. Substituting these expressions into \Eq{mutVar}, the balance between selection and mutation occurs when the allele frequency is
\begin{equation}\label{eq:haldaneDom}
  q = \frac{\Gm}{s}.
\end{equation}
This result applies to haploid genetic systems, and at least approximately to diploid systems with dominant deleterious mutations under the commonly used assumptions in population genetics. This expression captures the essential opposition between selective improvement and transmission decay that plays a key role in many biological problems. For the following section, it will be useful to note that, from \Eq{sV2}, the selection bias is $\GD S = -sV_q=-sq(1-q)$.

To complete the classic treatment, I now write the case for a recessive mutation in diploid genetics.   The mutation bias remains $\GD\Gt = \Gm(1-q)$.  For recessive alleles, the deleterious phenotype is only expressed in the homozygote, which occurs at frequency $q^2$ under random mating.  Thus, selective intensity on each copy of the deleterious allele increases with the probability, $q$, that it will be mated with another deleterious allele, so the selective intensity is $s_q=sq$.  Substituting these expressions into \Eq{mutVar} yields the classic mutation-selection balance for recessive diploid genetics as
\begin{equation}\label{eq:haldaneRec}
  q = \sqrt{\frac{\Gm}{s}}.
\end{equation}
For the following section, it will be useful to note that the selection bias against the deleterious allele is $\GD S = -sqV_q=-sq^2(1-q)$.

\begin{figure}[t]
\centering
\includegraphics[width=3in]{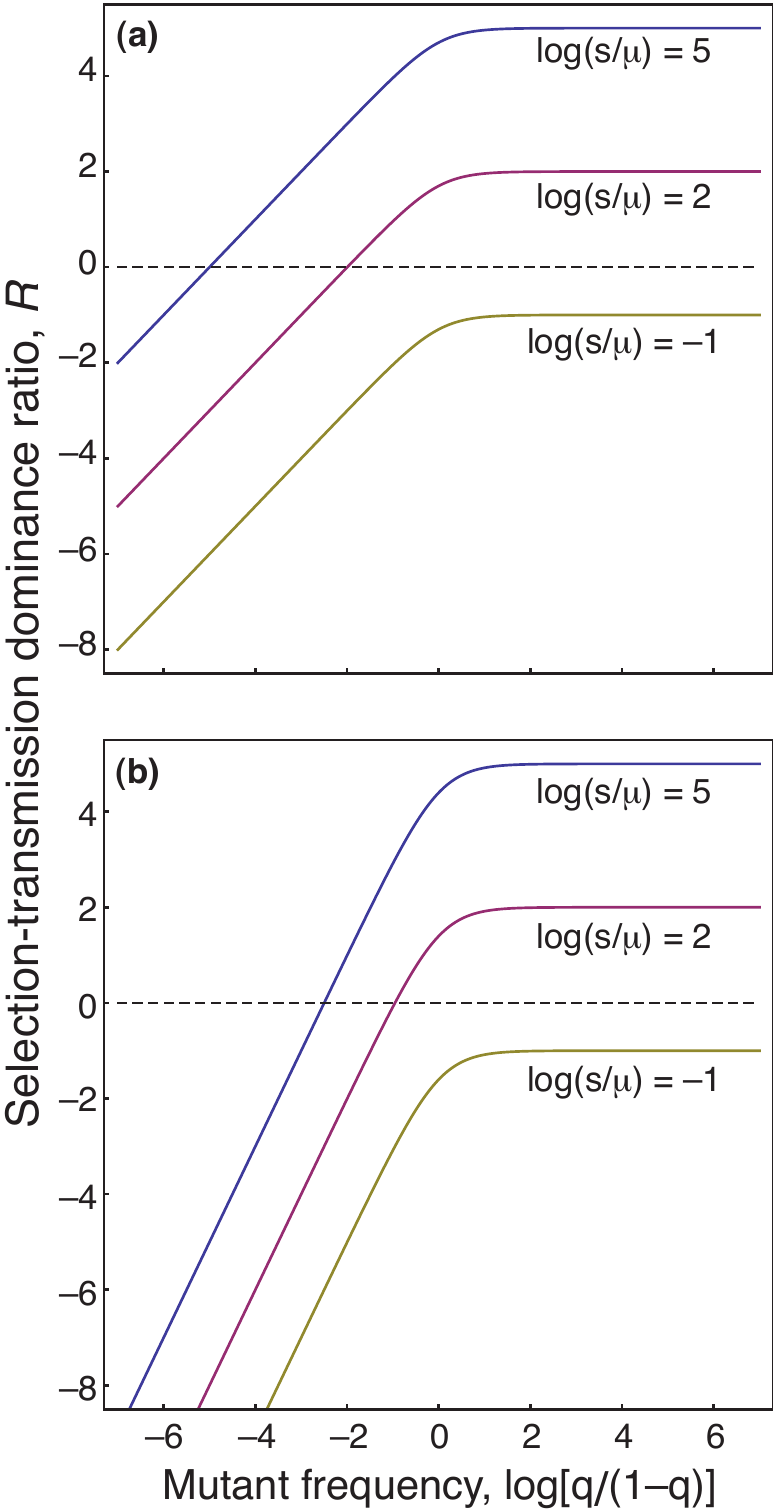}
\caption{The relative dominance of selection bias versus transmission bias in models of selection and mutation.  Relative dominance is measured by the ratio, $\R$, of \Eq{dominance}.  (a) The diploid dominant or haploid model.  (b)  The diploid recessive model, in which $\GD S = -sq^2(1-q)$ and $\R=\log(sq^2/\Gm)$.  All logarithms use base 10.
\label{fig:mutation}
}
\end{figure}

\subsection*{Ratio of selection versus transmission}

The epigraph from \textcite{williams66adaptation} emphasizes the relative strength of selection bias to transmission bias.  That comparison makes sense intuitively.  However, when we use the results in this section to measure the relative strength of selection and transmission, the comparison turns out to be complex.  The problem is that the relative strength of selection and transmission changes as evolution occurs in response to those forces.

For the simple models of selection and mutation in this section, \Fig{mutation} plots the relative strength of selection bias to transmission bias, $\R$, from \Eq{dominance}.  For example, \Fig{mutation}a shows the first model with equilibrium $q= \Gm/s$ in \Eq{haldaneDom}.  In that case, $\GD S  =-sq(1-q)$ and $\GD\Gt= \Gm(1-q)$, so the ratio is $\R=\log(sq/\Gm)$.   

The top curve of \Fig{mutation}a plots $\R$ for $\log(s/\Gm) = 5$.  The plot scales the frequency of the mutant allele as $\log[q/(1-q)]$. That scaling puts the midpoint of zero at $q=0.5$, with high and low frequencies scaling symmetrically and roughly logarithmically about the midpoint. 

In the top curve of \Fig{mutation}a, when the mutant frequency is not too low, selection bias is many times the transmission bias, $\R\gg0$.  However, the mutant frequency evolves in response to the relative strength of selection and transmission.  When selection is stronger, the mutant frequency, $q$, declines.  As $q$ declines, the ratio drops until $\R=0$, at which point the selection bias equals the transmission bias.  Similarly, when $q$ is very small, then the transmission bias is much greater than the selection bias, $\R\ll 0$, and the mutant frequency increases until the point $\R=0$.  

The ratio of the selection bias to the transmission bias does not have a constant value.  As mutant frequency changes, the relative dominance of the two forces shifts.  The system comes to rest only when selection and transmission are in balance.  Given the changing relation of selection and transmission, how should we interpret Williams' dictum?  

We could emphasize the example of the lower curve in \Fig{mutation}a.  That curve never rises above zero, because transmission bias is always greater than selection bias for all frequencies.  In that case, no hereditary information accumulates.  So we might say that hereditary information accumulates when selection bias is stronger than transmission bias for at least some conditions.  But that is a rather weak statement, changing Williams apparently beautiful clarity into a muddle.

Let us hold the point for now.  As we go through various examples, we will see that the ratio of selection bias to transmission bias changes in response to key aspects of the particular problem under study.  Rather than trying to abstract away how each particular problem shapes the changing ratio between selection and transmission, it may be more useful to use that ratio to understand each particular problem and the relations between different problems.

\subsection*{Timescale}

Timescale arises implicitly in these models, because selection and transmission are both expressed per unit time.  In the simplest models, one usually considers a single round of mutation per generation for each round of selection per generation.  However, multiple rounds of mutation can occur for each round of selection.  For example, many replications typically occur in the male germline of species that make large numbers of sperm.  Those multiple replications occur for each round of selection.  The multiple replications apparently increase the mutation rate in relation to the strength of selection \autocite{nachman00estimate}.  This change in the relative magnitudes of selection and mutation is important but not particularly profound.  Later, we will see more interesting ways in which timescale alters the balance between selection bias and transmission bias.

\section*{Variance under a balance between mutation and stabilizing selection}

Selection sometimes acts in a stabilizing way, pushing the average phenotype toward an intermediate optimum.  Mutation opposes selection by spreading trait values and increasing the average distance from the optimum.  The decrease in phenotypic variance caused by selection is opposed by the increase in variance caused by mutation. 

Here, I assume that all phenotypic variance is caused by simple genetics. This assumption allows me to focus on the processes that balance selection and mutation.  I summarize the standard approach for this problem \autocite{lande75the-maintenance,turelli84heritable,barton87adaptive}, following \textcite{frank90the-distribution}.

\subsection*{General expressions}

Define the character of interest as $\Gg=z^2$, and set the optimum at zero, which is also the average value in this symmetric model.  Then $z^2$ is the squared distance from the optimum, and the average of this squared distance is the variance.  Using $\Gg$ as the character of interest, at mutation-selection balance, from Eqs.~(\ref{eq:varianceBalance}) and (\ref{eq:mutVar}) we have 
\begin{equation}\label{eq:mutStableVar}
	V_\Gg = \frac{\GD\Gt}{s_\Gg}.
\end{equation}

Suppose a mutation adds or subtracts $c$ from the phenotypic value, $z$. The two directions of change occur with equal probability.  Thus, each mutation changes phenotype by $\pm c$. The contribution of each mutation to the change in squared deviation of phenotype, $z^2$, is, on average, $c^2$. Mutations happen at a rate $\Gm$, so the change in the phenotypic variance caused by mutation is 
\begin{equation*}
	\GD\Gt=c^2\Gm.
\end{equation*}

The scaling $c^2$ translates between genetic mutations and phenotypic effects.  We can use that same scaling to translate between the phenotypic scale, $\Gg$, and the genetic scale, $\Ga$, with the relation $\Gg = c^2\Ga$.  Here, $\Ga$ is the squared deviation on the genetic scale, and $\Gg$ is the squared deviation on the phenotypic scale.  The average of squared deviations is the variance, so we have $\overline{\Gg}$ and $\overline{\Ga}$ for the phenotypic and genetic variances, where the overbar denotes the average.

The term $V_\Gg$ is the variance of the squared phenotypic deviations, $\Gg$.  Because a variance is itself a squared value, $V_\Gg$ summarizes the square of the squared deviations, thus scaled to the fourth power.  Therefore, the proper relation to go from the phenotypic scale to the genetic scale is $V_\Gg = c^4V_\Ga$.  

Substituting $V_\Gg = c^4V_\Ga$ and $\GD\Gt=c^2\Gm$ into \Eq{mutStableVar} yields
\begin{equation}\label{eq:mutvarGeneral}
	V_\Ga=\frac{\Gm}{s},
\end{equation}
where $s=c^2s_\Gg$.  This expression for $V_\Ga$ provides the most general solution for variation under a balance between mutation and stabilizing selection.  However, $V_\Ga$ is the variance of squared deviations
 \begin{equation*}
	V_\Ga=\overline{\Ga^2}-\overline{\Ga}^2,
\end{equation*}
and thus scales with the fourth power of deviations.  Typically, we seek expressions for the variance under stabilizing selection rather than expressions scaled to the fourth power of deviations.  We can, under two particular cases, reduce the fourth power expression to an expression for variance under stabilizing selection.

\subsection*{Equilibrium variance}

When selection is much stronger than mutation, $s\gg\Gm$, the general balance result of \Eq{mutvarGeneral} is approximately
\begin{equation}\label{eq:varStrong}
	\overline{\Ga}\approx\frac{\Gm}{s},
\end{equation}
where $\overline{\Ga}$ is the variance on the genetic scale.  Note that this result is essentially the same as the haploid mutation-selection balance result in \Eq{haldaneDom} from the previous section.  \Boxx{variance} provides the derivation.

When selection is much weaker than mutation, $s\ll\Gm$, then 
\begin{equation}\label{eq:varWeak}
	\overline{\Ga}\approx\sqrt{\frac{\hat{\Gm}}{s}},
\end{equation}
which matches the result for the diploid recessive model in \Eq{haldaneRec}.  Here, $\hat{\Gm} = \Gm/2$.  With weak selection, most alleles deviate from the optimum of zero.  At nonzero values, mutation is equally likely to move the allelic value closer or farther from the optimum.  Thus, only one-half of mutations are deleterious, and $\hat{\Gm}$ expresses the deleterious mutation rate.  \Boxx{variance} provides the derivation.

Note that selection on phenotypes can be strong, yet the selection bias against each mutational step can be weak.  Here, \textit{weak selection} refers to the effect on each mutational step.  In particular, I defined $s=c^2s_\Gg$ below \Eq{mutvarGeneral}.  If the phenotypic effect, $c$, of each mutation is small, then strong selection on the phenotypic scale, $s_\Gg$, can be associated with weak selection on each mutational step of size $c$ when expressed on the genetic scale, $s$.

\subsection*{Mutation overwhelms selection}

If the decay in fitness by mutation exceeds the maximum fitness that can be achieved, then mutation overwhelms selection.  Mutation dominates selection when the magnitude of mutational effects is much greater than magnitude of selection, $s\ll\Gm$, which corresponds to results above for weak selection.  

\begin{figure}[H]
\begin{minipage}{\hsize}
\parindent=15pt
\noterule
{\bf \noindent\BoxLabel. Variance under mutation-selection balance}
\noterule
To obtain the equilibrium genetic variance in \Eq{varStrong} when selection is much stronger than mutation, $s\gg\Gm$, note that $\Gg=c^2\Ga$.  Thus, with strong selection, most alleles will be at the optimum with $\Gg=\Ga=0$, and a few alleles will be one mutational step away from the optimum at $\Gg=c^2$ and $\Ga=1$ \autocite{frank90the-distribution}.  Let the mutant frequency be $q$, so that $\Ga=1$ with probability $q$, therefore $\Ga^2=1$ with probability $q$.  Thus $\overline{\Ga}=\overline{\Ga^2}=q$ and $\overline{\Ga}^2=q^2$. With small $q$, we have $q\gg q^2$, therefore $V_\Ga=\overline{\Ga^2}-\overline{\Ga}^2\approx \overline{\Ga^2}= \overline{\Ga}$, and thus \Eq{mutvarGeneral} leads to \Eq{varStrong}.

To obtain the equilibrium genetic variance in \Eq{varWeak} when selection is much weaker than mutation, $s\ll\Gm$, we assume that the distribution of allelic values approximately follows a Gaussian with a mean at zero \autocite{kimura65a-stochastic,lande75the-maintenance,frank90the-distribution}.  With a Gaussian, the fourth moment is approximately three times the square of the second moment (variance), thus $\overline{\Ga^2} \approx 3\overline{\Ga}^2$ and $V_\Ga=\overline{\Ga^2}-\overline{\Ga}^2\approx 2\overline{\Ga}^2$.  Using this expression for $V_\Ga$ in \Eq{mutvarGeneral} yields \Eq{varWeak}. 
\noterule
\end{minipage}
\end{figure}
\boxlabel{variance}

From \Eq{fitRegress} of \Boxx{price}, we can write fitness as $w = 1 - s\Ga$, using $s\equiv s_\Ga$ for selective intensity on the genetic character $\Ga$.  Thus, average fitness is $\bar{w}=1-s\overline{\Ga}$ and, using \Eq{varWeak} for $\overline{\Ga}$, we obtain $\bar{w}=1-s\sqrt{\hat{\Gm}/s} = 1-\sqrt{\hat{\Gm} s}$.  Mutational meltdown occurs when $\bar{w} < 0$, which implies $\hat{\Gm} s > 1$.

This condition simply means that the amount of deleterious mutation, $\hat{\Gm}$, scaled by the fitness consequence per mutation, $s$, reduces fitness by an amount that is greater than maximal fitness.  The next section considers when the mutation rate might be so high.

\section*{Error threshold and quasispecies}

Eigen applied the fundamental tension between mutation and selection to the evolution of nucleotide sequences.  In early evolution, the mutation rate was likely to be high because enzymes that correct replication errors did not yet exist.  The initially high mutation rate and lack of error correction lead to Eigen's error threshold paradox \autocite{eigen71self-organization,eigen92steps,eigen77the-hypercycle.,maynard-smith79hypercycles}.  

Suppose the initial replicating sequences had a length of $n$ nucleotides.  If the mutation rate per nucleotide is $\Gm$, then the mutation rate per sequence is roughly $n\Gm$.  The deleterious effect per mutation is $s$.  Thus, the expected deleterious effect of mutation during each replication of a sequence of length $n$ is $n\Gm s$.  When the deleterious effect per replication is greater than maximum fitness, here scaled to be one, mutation overwhelms selection and no selective increase in adaptation can be achieved.  The condition for remaining below this error threshold is $n\Gm s<1$, which means that sequence length is limited to 
\begin{equation*}
  n < \frac{1}{\Gm s}.
\end{equation*}

Eigen noted the paradox of the error threshold for early evolution.  Without error correcting enzymes, the mutation rate was high.  A high mutation rate limited the maximum sequence length.  A short sequence could not contain enough information to encode error correcting enzymes.  Without error correcting enzymes, the sequence remains too short to encode error correction.  How did the biochemical machinery of error correction evolve?

\textcite{eigen88molecular,eigen89the-molecular} discussed a second interesting property of sequence evolution under mutation and selection.  A population of sequences exists as a mixture of the most fit sequence and a variety of mutant sequences.  Eigen called the most fit sequence the \textit{master sequence}, and the population of sequences that are zero, one, two, or more mutational steps away from the master sequence the \textit{quasispecies}.  The term \textit{quasispecies} is meant to differentiate a population of variants from a typological notion of a species as a fixed, nonvarying entity.

The error threshold and the quasispecies are equivalent to the standard evolutionary concepts of heritable variation maintained by a balance between mutation and selection, as described in the previous section \autocite{wilke05quasispecies}.  The epigraph from Williams captured the key idea of the error threshold by expressing the notion of a gene ``as any hereditary information for which there is a $\ldots$ selection bias equal to several or many times its rate of endogenous change'' \autocite[p.~44]{williams66adaptation}. The classical mutation-selection theory of Haldane, extended to the maintenance of variation under stabilizing selection, expresses the concept of quasispecies.  All of these theories have to do with the fundamental partition of total evolutionary change into a component of selection and a component of transmission fidelity.

\section*{Multilevel analysis of clade selection}

\textcite{williams92natural} argued that the relative rates of selection and transmission influence evolutionary change at all taxonomic levels.  Williams adopted the term \emph{clade selection} from \textcite{stearns86natural}. Van Valen's \citeyear{van-valen75group} analysis provides the clearest way to understand the ideas and potential importance.

\textcite{van-valen75group} began by comparing the evolutionary history of sexual and asexual types.  He set up the problem by assuming that asexuals have a short-term advantage in growth rate relative to sexuals, and that sexuals have a long-term advantage with regard to avoiding extinction and forming new species \autocite{fisher30the-genetical,stebbins50variation}. With those assumptions, Van Valen (p.~87) suggests that one
\begin{quote}
Consider a large set of species, some obligatory apomicts [asexuals] and some at least facultatively sexual. 
The apomicts will have a greater probability of extinction of lineages and the sexual species will have a greater probability of speciation by splitting of lineages. $\ldots$ However, apomicts will sometimes originate from sexual species because of their immediate advantage.
\end{quote}

Van Valen recognized two levels of selection. Clades with more sexual species will increase in species number relative to clades with fewer sexual species. Thus, sex has an advantage between clades. Within clades, asexuals will arise repeatedly because of their short-term advantage relative to their sexual ancestor. The selection within clades that favors asexuals can be thought of as a transmission bias: sexual species sometimes produce asexual descendants, whereas asexual species rarely produce sexual descendants.

Van Valen used the fact that one can express the two levels of selection as selection between clades and a transmission bias within clades to develop a simple model for the equilibrium frequency of asexuals.  That equilibrium frequency balances selection bias between clades favoring sexuals, with rate $s$, against transmission bias within clades favoring asexuals, with rate $\Gm$, to obtain the approximate equilibrium frequency of asexuals, $q$, as
\begin{equation*}
	q \approx \frac{\Gm}{s}.
\end{equation*}
This expression is the same as the standard model of mutation-selection balance in genetics given in \Eq{haldaneDom}.  In this model, \textcite{van-valen75group} emphasizes that selection at any taxonomic level is always potentially balanced against the rate of endogenous change at that level, echoing the epigraph from Williams.  Endogenous change may arise in various ways, such as mutation by change of state or selection between the lower-level entities that comprise the higher level.

Van Valen also applied this approach to mammals.  In mammals, genera with larger body size survive longer than genera with smaller body size, but the smaller bodied genera bud off new genera at a higher rate.  The net reproductive rate of small genera is higher, giving a selective advantage to small bodied genera over large bodied genera. Within genera, there is a bias toward larger body size. The distribution of mammalian body size is influenced by the balance between selection between genera favoring smaller size and selection within genera favoring larger size.

Various philosophical issues in the interpretation of clades as units have been taken up by \textcite{valen88species}, \textcite{williams92natural}, and \textcite{okasha06evolution}. Here, I only applied the fact that one can partition the patterns of change at one level, such as clades, into components of selection and transmission. The philosophical issues focus on whether one can think of clades as natural units, for some reasonable meaning of \emph{natural.}

\section*{Multilevel analysis of kin and group selection}

Total evolutionary change includes a part caused by selection and a part caused by lack of fidelity in transmission (Eq.~\ref{eq:total}). In this section, I use that basic partition of total change to study two levels of selection, generalizing the model of clade selection in the prior section.  

At the higher level, a group may be any sort of collection.  We may, for example, consider groups of individuals or groups of alleles within an individual.  Selection concerns differential success among groups.  At the lower level, selection within groups causes a bias in the transmission fidelity of group-level characteristics \autocite{lewontin70the-units,price72extension,hamilton75innate,wilson89reviving,okasha06evolution}. 

The most interesting problems arise when selection among groups opposes the transmission bias caused by selection within groups.  We may then consider a balance between selection and transmission or, equivalently, a balance between the two levels of selection, $\GD S = -\GD\Gt$, as in \Eq{totalBalance}.  

I present three aspects of multilevel selection.  First, I write a very simple expression for the balance between the two levels of selection.  This expression of balance provides the general basis for multilevel models of selection and the analogy to the classical models of selection and mutation.

Second, I apply the balance between different levels of selection to the tension between competition and cooperation.  That simple model illustrates how easily we can understand the basic processes of group-level cooperation within the broader framework of selection and transmission.  I also show the fundamental equivalence of group selection and kin selection models in group-structured populations.

Third, I discuss the roles of population regulation and timescale.  For population regulation, if limited space or resources regulates group productivity, then all groups may have roughly the same reproductive output.  In that case, little selection occurs among groups, and the within group component of selection dominates \autocite{wade85soft,frank86hierarchical,frank98foundations,taylor92altruism,wilson92can-altruism,queller94genetic}.  For timescale, the number of rounds of selection within groups relative to the rate of selection among groups sets the relative scaling of selection between the two levels.  When the rate of selection within groups overwhelms the rate of selection among groups, then the within group component of selection dominates evolutionary process \autocite{williams66adaptation}.

\subsection*{The balance between levels of selection}

The fundamental equation for balance is $\GD S = -\GD\Gt$, the balance between selection bias and transmission bias.  For multilevel selection, we interpret $\GD S$ as the selection among groups and $\GD\Gt$ as the transmission bias caused by selection within groups.  For problems in which selection at the different levels pushes character values in opposing directions, we may rewrite the expression as $\GD S_a = -\GD S_w$, the balance between selection among groups and selection within groups.

The change in a character caused by selection can be expressed as $\GD S = sV$, the product of the selective intensity, $s$, and the variance in the character under selection, $V$ (see \Boxx{price}).  Thus, we may write the balance $\GD S_a = -\GD S_w$ as
\begin{equation*}
 s_aV_a = -s_wV_w.
\end{equation*}
In a group-structured population, the total variance is the sum of the variance among groups and the variance within groups, which we express as $V_t=V_a+V_w$. Making the substitution $V_w=V_t-V_a$ yields
\begin{equation*}
 s_aV_a = -s_w(V_t-V_a).
\end{equation*}
It is convenient to express the pattern of variance by the correlation coefficient $r=V_a/V_t$, where $r$ measures the correlation in character values between individuals within a group.  Dividing both sides by $V_t$ yields
\begin{equation}\label{eq:multi}
 s_a r = -s_w(1-r).
\end{equation}
To understand this expression, we need to consider the interpretation of the correlation, $r=V_a/V_t$. The correlation is the fraction of the total variance that is among groups. Because variance provides a weighting on selection, $r$ can be thought of as the fraction of the total weighting of selection that happens at the group level, and $1-r$ can be thought of as the fraction of the total weighting of selection that happens within groups.  

Thus, $s_a r$ is the intensity of selection among groups, $s_a$, multiplied by the weighting of selection at the group level, $r$.  At a balance, the group-level component must be equal and opposite to the intensity of selection within groups, $s_w$, multiplied by the weighting of selection within groups, $1-r$.  

The correlation $r$ is also a particular form of the regression coefficient of relatedness from kin selection theory, as hinted initially by \textcite{hamilton75innate,hamilton79wingless} following from the work of \textcite{price72extension}, and later analyzed more formally \autocite{grafen84natural,wade85soft,frank86hierarchical,frank98foundations,west07social}. The equivalence of $r$ and Hamilton's formal theory of kin selection establishes the exact equivalence of multilevel group selection and kin selection.

\subsection*{The tension between competition and cooperation}

We need an explicit expression for the relation between a trait and fitness in order to evaluate the abstract expressions from the previous section.  In this section, I present a simple model of competition and cooperation \autocite{frank94kin-selection,frank95mutual}.

In a group-structured population, we can express fitness as the product of two components.  The first component is the individual's relative share of total group success.  The second component is the total success of the group.  For the first component, we may write the individual's relative share of the group's success as $z/z_g$, where $z$ is the individual's tendency to be competitive against neighboring group members for access to local resources, and $z_g$ is the average competitive tendency in the individual's group.  Selection within groups always favors greater competitive tendency, because an individual's share of group success always rises with an increase in $z$.

For the second component, total group success, suppose that the more intensely individuals compete against neighbors, the less efficient the group is in using its resources productively.  For example, a certain fraction of local energy may go into outcompeting neighbors rather than enhancing productivity.  We may express the negative effect of competitiveness on group productivity by writing the total group productivity as $1-z_g$, in which the total productivity declines as the group members' average tendency to compete, $z_g$, rises.  Thus, selection among groups always favors a less competitive and more cooperative behavioral tendency, because group success declines as average competitiveness, $z_g$, rises.

Putting the two pieces together, the fitness, $w$, of an individual with competitive tendency, $z$, in a group with average competitive tendency, $z_g$, is
\begin{equation}\label{eq:indFit}
 w = \frac{z}{z_g}\left(1-z_g\right).
\end{equation}

To evaluate the balance between selection at the group level and selection within groups, we need to relate the expression for fitness to the particular selective tendencies and variance components in \Eq{multi}.  The selective intensity among groups is $s_a=-1$, because group fitness is $1-z_g$, and selective intensity is the change (partial derivative) in group fitness with change in the average trait value in the group.  The selective intensity within groups is $s_w=\left(1-z_g\right)/z_g$, which is the change in individual fitness, $w$, with change in individual character value, $z$.

Substituting these values for $s_a$ and $s_w$ into \Eq{multi} yields a balance between group and individual selection when
\begin{equation}\label{eq:tragedyBalance}
 -r=-\frac{1-z_g}{z_g}(1-r).
\end{equation}
Skipping over the technical details, we may say roughly that, in this case, selection acts in a stabilizing way, causing individuals trait values to converge toward a single value that is an evolutionarily stable strategy (ESS).  Thus, individual values, $z$, converge toward group averages, $z_g$, which in turn converge to a global value, $z^*$.  Making the substitution $z_g=z^*$ and solving for $z^*$ gives the balance point \autocite{frank94kin-selection,frank95mutual} as
\begin{equation}\label{eq:tragedySoln}
 z^* = 1-r.
\end{equation}

This balance point expresses the key insights of multilevel selection and kin selection.  In terms of multilevel selection, $1-r$ is the fraction of the total variance that occurs within groups.  The greater this weighting of within-group selection, the higher the balancing point of $z^*$, the tendency of individuals to compete with neighbors.  As variance shifts toward the group level, $1-r$ declines, $z^*$ decreases because competitive restraint is more strongly favored, and the balance of selective forces increasingly favors cooperative behavior.  In terms of kin selection, as the coefficient of relatedness, $r$, increases, competitive restraint and cooperative behavior rise.

\subsection*{Population regulation and timescale}  

Several factors may influence the intensity of selection within groups compared with the intensity of selection among groups \autocite{alexander78group,wade85soft}.  For example, if limited space or resources regulates group productivity, then all groups may have roughly the same reproductive output.  In that case, little selection occurs among groups, and the within group component of selection dominates \autocite{wade85soft,frank86hierarchical,frank98foundations,taylor92altruism,wilson92can-altruism,queller94genetic}

In the model from the prior section, group productivity was $1-z_g$, and the change in group productivity with a change in the group phenotype, $z_g$, was $s_a=-1$.  Suppose instead that the relation between group phenotype and group productivity is much weaker, because extrinsic aspects of space and resources limit group productivity.  For example, if group productivity is $1-\Ge z_g$, where $\Ge<1$, then $s_a=-\Ge$.  Using that value of $s_a$ in \Eq{tragedyBalance}, we obtain the solution
\begin{equation*}
 z^* = \frac{1-r}{1-r(1-\Ge)}
\end{equation*}
This solution is equivalent to \Eq{tragedySoln} when $\Ge=1$.  As limits on group productivity become more stringent, $\Ge$ declines toward zero, the balance tips more strongly in favor of selection within groups, the level of competitiveness, $z^*$, increases, and, equivalently, the level of cooperation declines.  

There are, of course, many complex ways in which individual traits may relate to group productivity and to the intensity of selection within groups.  But all the different complexities tend to reduce to the simple balancing of forces between selection among groups versus bias in transmission fidelity of group characteristics or, equivalently, selection among groups versus selection within groups.  If some additional force weakens selection among groups, then selection within groups increasingly dominates.  Similarly, if some additional force weakens selection within groups, then selection among groups increasingly dominates.

Timescale provides another example.  If the number of rounds of selection within groups increases relative to the pace of selection among groups, then selection within groups increasingly dominates the balance of forces \autocite{frank86hierarchical,frank87demography}.  I discuss two particular cases in later sections on parasite virulence and microbial metabolism.

\begin{figure}[t]
\centering
\includegraphics[width=3in]{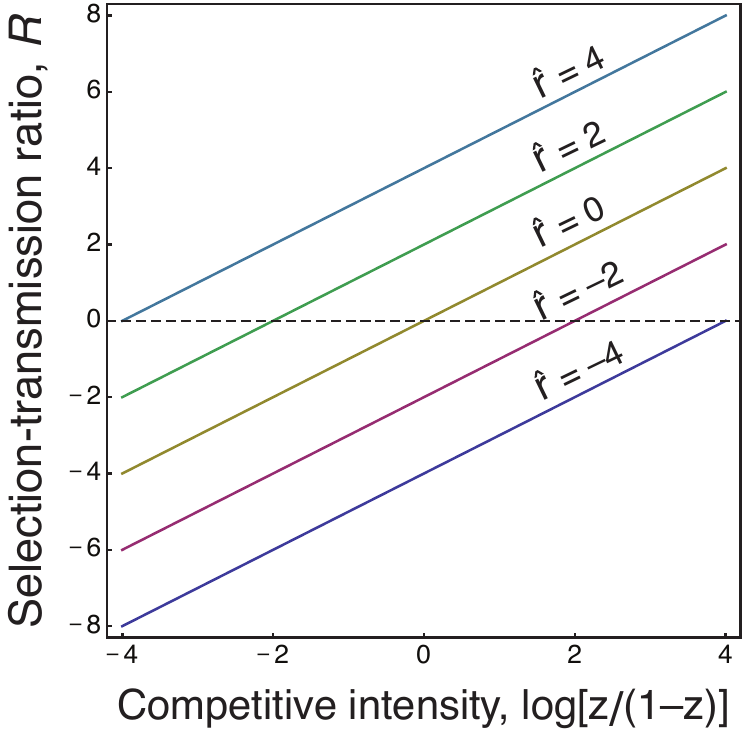}
\caption{The relative dominance of selection bias among groups versus transmission bias within groups in a multilevel selection model.  Relative dominance is measured by the ratio, $\R$, of \Eq{multiDom}.  Different levels of relatedness shift the balance between selection bias among groups and transmission bias within groups.  Here, relatedness is measured by $\hat{r}=\log[r/(1-r)]$.
\label{fig:multilevel}
}
\end{figure}

\subsection*{Ratio of selection at different levels}

In multilevel selection, the relative strength of selective bias to transmission bias from \Eq{dominance} compares selection among groups to selection within groups.  Substituting the expressions for $\GD S$ and $\GD\Gt$ derived from \Eq{indFit} into \Eq{dominance} yields

\begin{equation}\label{eq:multiDom}
	\R = \log\left(\frac{z}{1-z}\right) +\log\left(\frac{r}{1-r}\right).
\end{equation}
\Fig{multilevel} plots $\R$ versus the level of competitiveness, $z$, for different levels of relatedness, $r$.  

The level of competitiveness is in equilibrium balance, $z^*$, when the lines cross $\R=0$. At that point selection bias among groups is equal and opposite to transmission bias caused by selection within groups.  Once again, we see that selective bias is greater than transmission bias only when the system is out of equilibrium.  

Following the epigraph from Williams, one may wish to think of groups as acquiring information, adaptation, or a degree of unitary function to the extent that selective bias tends to dominate transmission bias.  Because relative dominance depends on the phenotype, $z$,  one interpretation would be that significant group-level function requires the relative dominance of selection over transmission across a wide range of possible phenotypes \autocite{gardner09capturing}.  The range of phenotypes over which selection bias dominates transmission bias increases with a rise in relatedness, $r$. Thus, one may say that increasing relatedness shifts the locus of information or adaptation toward the higher level.  

That interpretation of group-level unity or adaptation goes beyond what the analysis by itself presents.  The analysis simply describes the way in which $\R$ shifts with competitive intensity and relatedness. The interpretation of group-level unity is a gloss that may aid or hinder understanding in different contexts.  Ultimately, one must retain a clear view of the underlying analytical basis.

\section*{Stochastic corrector model of early protocells}

Protocells are simple membrane-bound groups of genes that likely formed in early evolution \autocite{maynard-smith95the-major}.  A model of protocell evolution provides insight into group selection, kin selection, parasite virulence, and the evolution of symbionts \autocite{szathmary87group,frank94kin-selection,frank96models,frank97models}.  

In the protocell model, the selective bias between cells opposes the transmission bias arising from selection between genes within cells.  Expanding on the epigraph from \textcite{williams66adaptation}, the degree to which adaptive design occurs at the protocell level versus the internal genic level depends on the selective bias between cells relative to the rate of endogenous change within cells.

Each protocell can be thought of as a bag that starts with $k$ pieces of genetic material (chromosomes).  The chromosomes compete within the protocell for resources. Success at acquiring resources influences the rate at which chromosomes can replicate themselves within the cell. More competitive chromosomes use up local resources less efficiently and reduce the overall success of the protocell and its group of chromosomes. 

A protocell competes with other protocells for resources from the environment. A protocell produces a progeny cell after it has acquired sufficient resources and its chromosomes have replicated. The fitness of the protocell and its chromosomes depends on the rate of progeny production. Sampling of chromosomes occurs when progeny are formed: $k$ chromosomes are chosen randomly from the pool of copies in the cell. I refer to this sampling process as segregation.

This protocell model is a particular expression of the group selection model in the previous section. By studying this particular example, we can see more clearly how specific aspects of mutation, competition, and selection within groups affect transmission bias.

Suppose that the fitness of a chromosome follows the expression in \Eq{indFit}, repeated here
\begin{equation*}
 w = \frac{z}{z_g}\left(1-z_g\right),
\end{equation*}
where $z$ is a chromosome's tendency to be competitive against neighboring chromosomes for access to local resources within the protocell, and $z_g$ is the average competitive tendency of chromosomes in the protocell.  Following \Eq{tragedySoln} of the previous section, the balance of selection between protocells and transmission bias within protocells is $z^*=1-r$, where $r$ is the kin selection coefficient of relatedness among the chromosomes within a cell. 

\subsection*{Virulence and symbiosis}

The stochastic corrector model allows us to connect the abstract expressions from the multilevel analysis of kin and group selection to specific interpretations of parasite virulence and the evolution of symbionts within hosts \autocite{frank94kin-selection,frank96models}.  For virulence, one can think of each of the $k$ chromosomes as a parasite, and one can think of the protocell as the host.  Competition between the parasites may cause inefficient use of host resources.  Overexploitation of the host reduces host fitness.  Thus, competition between parasites within hosts tends to increase virulence.  The lower the relatedness, $r$, among the parasites within a host, the greater the competitiveness and virulence of the parasites, $z^*=1-r$. 

We may also think of the $k$ chromosomes as symbionts living within a host.  From the host's point of view, increasing $r$ reduces the competitiveness between the symbionts, aligning symbiont and host interests. In order to increase $r$, hosts may be favored to reduce the number, $k$, of symbionts transmitted to offspring or transmitted between hosts \autocite{frank96host}.  Hosts may also be favored to reduce the mixing of symbionts between different hosts \autocite{frank96host-symbiont}.

\subsection*{Kin selection and group selection}

This model allows us to evaluate the meaning of the kin selection coefficient, $r$, within a particular scenario.  Assume that transmission is purely vertical, because the chromosomes do not mix between cells.  In this model of vertical transmission, three forces affect the evolution of competitiveness, $z^*=1-r$. 

First, selection between protocells favors reduced competitiveness of chromosomes within cells, leading to greater efficiency at the cellular level. Against that cellular level effect, the competition and selection between chromosomes within cells causes a transmission bias that favors increased competitiveness of chromosomes. 

Second, mutations reduce the similarity among chromosomes within hosts, thus reducing $r$. The force imposed by mutation is controlled by two parameters, the mutation rate, $\Gm$, and the change in character value caused by each mutation, $\Gd$. Each mutational event changes $z$ by $\pm\Gd$, where the alternative directions of change occur with equal probability. Thus, mutation by itself causes no transmission bias.

Third, segregation samples from the local chromosomes when the protocell reproduces. Each new progeny starts with $k$ chromosomal copies. When the cell reproduces, replicates of the local chromosomes are chosen stochastically according to relative fitness within the cell, $z/z_g$. This sampling reduces the variance within hosts and increases relatedness.

A stochastic computer simulation of this model showed that relatedness, $r$, and equilibrium trait values, $z^*$, are held in balance by a delicate interaction among mutation, selection and segregation \autocite{frank94kin-selection}. The observed equilibrium trait values in the computer simulation closely follow the prediction $z^*= 1-r$, where $r=V_a/V_t$ is calculated directly from the simulation by measuring the within-cell and total variances of trait values for the individual chromosomes in the population.  The specific parameters affect variances and equilibrium trait values as expected: relatedness declines and $z^*$ rises as the mutation rate, $\Gm$, or mutation step, $\Gd$, increases. An increase in the number of chromosomes per cell, $k$, causes an increase in competitiveness, $z$, because more copies reduce the variation among cells caused by sampling during segregation.

\subsection*{Kin selection arises from patterns of variance,  not genealogy}

The analysis in the previous section demonstrates that genealogy does not provide a sufficient explanation for the evolution of cooperative and competitive traits.  The genealogical closeness between chromosomes in a cell increases as $k$ declines.  That genealogical aspect explains some of the changes in competitiveness, $z^*$.  But, for a fixed genealogical scheme and a fixed mutation rate, the magnitude of the effect of each mutation, $\Gd$, can strongly influence the equilibrium value, $z^*$ \autocite{frank94kin-selection}.  Larger mutational effects raise the variance within groups relative to the variance among groups, causing a decline in $r$, an increase in the strength of selection within cells, and an increase in the equilibrium competitiveness, $z^*$.  

The theory of kin selection formulated by \textcite{hamilton70selfish} depends solely on patterns of variance and correlation, not on genealogy \autocite{frank98foundations}.  Genealogy is often closely associated with patterns of variance and correlation. The simple protocell model illustrates how the association between genealogy and the patterns of variance and correlation may break down.  When the association breaks down, the true causal processes of variance and correlation explain the outcome.  Since Hamilton's \citeyear{hamilton70selfish} work, no fundamentally derived theory of kin selection based on genealogy has existed.  However, it is often convenient to use the fact that genealogy typically associates with the underlying causal processes of variance and correlation.  That convenience has unfortunately confused many authors about the distinction between a convenient association and the fundamental theory and its history.

We  may recover the association between genealogy and causal process if mixing of chromosomes between cells occurs.  Such mixing often dominates mutation in determining the patterns of variance within and among groups.  In that case, genealogy may become the main force determining $r$ and the equilibrium level of competitiveness, $z^*$.  

In conclusion, the mutation rate and the size of mutational effects primarily influence the patterns of variance under some conditions, whereas the migration rate and genealogy primarily influence the patterns of variance under other conditions.  It is the patterns of variance and correlation that determine outcome.

\subsection*{Reasons to favor kin selection over group selection}

Kin and group selection follow the same partition of variance within and among groups.  A group selection analysis tends to emphasize the variance among groups and therefore the effect of selection at the group level.  A kin selection analysis tends to emphasize the correlation between members of the same group, measured by the kin selection coefficient.  The correlation within groups and the relative amount of variation among groups are simply alternative ways of expressing the partitioning of variances \autocite{frank86hierarchical}.

In more complicated biological problems, it often becomes difficult to express all of the selective forces in terms of relative variances among groups.  The problem is that patterns of interaction may differ with respect to different processes, such as mating, competition between certain individuals such as males, and competition between other individuals such as females.  In that sort of realistic scenario, it is far easier to trace pathways of causation through a series of partial correlations that can be interpreted as an extended form of kin selection analysis \autocite{frank86hierarchical,frank98foundations}.  In practice, it is rarely sensible to express such multiple pathways of causation by expressions of relative amounts of variance among groups, although such expressions may be possible mathematically.  For that reason, kin selection often becomes a more natural form of analysis for realistic biological problems, leading to a generalized path analysis framework.  

The present article is about the separation between selection and transmission rather than a general approach to pathways of causation.  \textcite{frank97the-price,frank98foundations} summarized the path analysis approach, although some readers may find those publications a bit technical.  I will return to the path analysis methods in a later article in this series.

\section*{Short-sighted parasite evolution}

Within-group competitiveness often evolves, even though competitiveness reduces the equilibrium fitness of all individuals.  The models in the previous sections provided examples.  In those models, the favored value of competitiveness was given in \Eq{tragedySoln} by $z^*=1-r$.  Competitiveness rises as relatedness between group members, $r$, declines.  The equilibrium fitness from \Eq{indFit} is $w=1-z^*=r$.  Thus, reduced relatedness in groups increases competitiveness and causes a decline in fitness for all individuals and all groups.

I mentioned one interpretation of this simple model in terms of parasite virulence.  Parasites may compete for resources within the host.  Greater competitiveness may lead to overexploitation of the host, harming the host and ultimately damaging or destroying the resource on which the parasites depend.  In that regard, reduced relatedness of parasites within hosts may lead to enhanced competition and greater virulence, where ``virulence'' means the degree of harm the parasites cause the host. 

\textcite{levin94short-sighted} emphasized the key role of evolutionary timescale.  A long period of within-host evolution, with many rounds of parasite competition and selection, may favor the origin and spread of increasing competitiveness between parasites, leading to greater virulence.  That evolution of increasing virulence occurs during the time of an infection within a single host.  Such evolutionary increase of virulence can kill the host and, in consequence, kill the parasites themselves.  In that regard, the newly evolved virulence is short-sighted, because it provides a local advantage to the parasites in the short run but leads to their extinction in the long run.  

If the highly virulent forms that evolve within the host rarely transmit to other hosts, then two distinct timescales exist.  On the short timescale within hosts, high virulence repeatedly evolves but does not contribute to the long run evolution of the population.  On the long timescale in the population of parasites across hosts, the less virulent forms transmit between hosts better than do the highly virulent forms, causing a moderate to low level of virulence among infective parasites entering a host.  

By contrast, if the highly virulent forms that evolve within hosts often transmit to other hosts, then the shorter and longer timescales interact.  The short-term evolutionary increase of competitiveness within the host contributes to a transmission bias on the longer timescale.  The contribution of the short-term increase in virulence within hosts to the longer timescale depends on the fraction of parasites transmitted between hosts that come from the later population within the host.  The next section provides an example.

\section*{Demography, timescale and microbial metabolism}

In this section, I consider groups that continuously produce transmissible forms. The longer the time for evolution within groups, the greater the transmission bias toward characters favored within groups.  For example, within-group selection often favors greater competitiveness against neighbors.  If many generations of selection occur within groups, the greater short-term pressure for competitiveness within groups ultimately increases the competitiveness across all groups. 

Microbial metabolism nicely illustrates aspects of timescale \autocite{frank10the-trade-off}. Extra energy devoted to resource acquisition speeds metabolic rate and competitive success against neighbors but reduces net efficiency and yield. Thus, the local benefit for rapid resource acquisition trades off against lower yield and reduced competitive success of a group against other groups \autocite{pfeiffer01cooperation}. 

Once again, we have a situation in which selection within groups favors more competitive traits, whereas selection between groups favors greater restraint and higher group productivity.  The balance between opposing forces ultimately depends on the relative selective bias between groups compared with the transmission bias caused by selection within groups.

\subsection*{An example}

Suppose that individual microbial colonies occur in separated patches.  Each patch lasts for a while but eventually disappears.  During a patch's lifespan, there is a continual flow of resources available to the microbes.  The microbes compete for the resources within the patch.  Competition occurs by rate of resource uptake.  Individuals that invest more energy in uptake outcompete neighbors for resources, but their net conversion of resource into reproduction is lower because they spend more on uptake rather than productivity.  Groups that have highly competitive strains, devoting much energy to competitive increases in resource uptake, have low net productivity.  

Colonies continuously send out migrants in proportion to group productivity.  Transmission bias occurs when the average competitive trait of migrants differs from the average trait value among those microbes that founded the colony.  The local processes of competition, selection, and production of migrants continue until colony extinction.  Colony formation and colony extinction set the global birth-death process.  

The overall scenario is roughly similar to a host-parasite situation, in which resource patches are like hosts, and parasites send out transmissible progeny continuously from an infected host.  Many variations are possible.  However, the basic setup provides a useful expression for the interactions between colony demography and the different timescales of selection bias and transmission bias.

I use the particular assumptions and results of \textcite{frank10the-trade-off}.  The interpretation follows the same type of selection-transmission balance of previous sections.  However, the earlier models were often designed explicitly to illustrate the partition between selective bias and transmission bias.  The value here arises from the more realistic biology, which forces us to parse the components of evolutionary change without the advantage of a toy model designed to give us a simple partition.

\begin{figure}[t]
\centering
\includegraphics[width=\hsize]{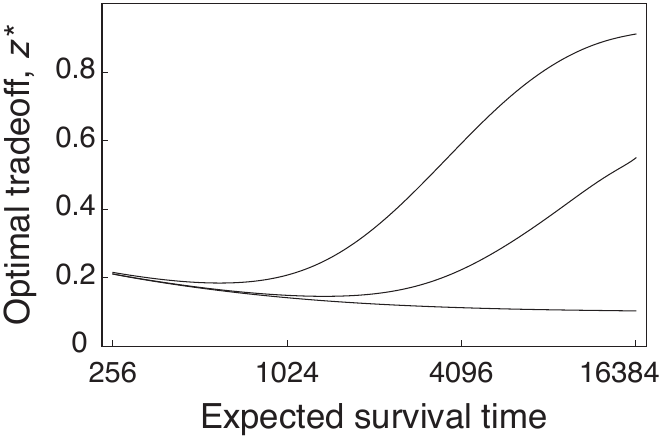}
\caption{
The trade-off between rate and yield in microbial metabolism. The optimal trade-off, $z^*$, is the fraction of available resources invested in increasing the rate of acquiring new resources.  The remainder of resources, $1-z^*$, enhances reproduction.  The colony survives each time period at rate $\Gd$; the expected survival time is $1/\Gd$. Each colony begins with a single immigrant or small group of genetically identical immigrants. The microbes use the local resources to reproduce.  Mutations occur in the trade-off, $z$, between rate and yield. The lower curve represents no mutation in the colony.  The middle curve has mutation rate, $\Gm$, and the upper curve has a higher mutation rate of $10\Gm$. The colony sends out migrants to colonize new patches.  The number of migrants per unit time for each genetic type in a patch is proportional to the number of cells of that genetic type. The details about rate processes are in \textcite{frank10the-trade-off}. Redrawn from Fig.~2a of \textcite{frank10the-trade-off}.
\label{fig:microbe}
}
\end{figure}

\subsection*{No transmission bias}

\Fig{microbe} shows the net outcome of selective bias between groups and transmission bias within groups.  Each colony forms by a small group of genetically identical cells.  When there is no mutation, as shown in the bottom curve, no selection can occur within the colony because there is no genetic variation.  Thus, the bottom curve reflects the pure effects of selective bias between groups in the absence of transmission bias.  The character, $z$, is the fraction of energy devoted to resource acquisition relative to the fraction $1-z$ devoted to reproduction.  The character value at which equilibrium occurs is $z^*$.

To understand the consequences of a pure selective bias between groups, recall from \Eq{total} that the total change in a character is
\begin{equation*}
  \wbar\GD\zbar=\GD S + \GD\Gt,
\end{equation*}
the sum of the change caused by selective bias, $\GD S$, and transmission bias, $\GD\Gt$.  Here, the biases are measured with respect to microbial groups living in isolated patches.

The character value settles to equilibrium when $\wbar\GD\zbar=\GD S + \GD\Gt=0$.    If there is no genetic variation among the initial microbes that start each colony, and no mutation, then there can be no selection within groups and no transmission bias, thus $\GD\Gt=0$.  With no transmission bias, the system comes to equilibrium when $\GD S =0$.  Put another way, group productivity, which determines the selective bias between groups, $\GD S$, sets the trade-off between rate and yield.  In the lower curve of \Fig{microbe} with no mutation, the value $z^*$ maximizes yield and leads to $\GD S = sV = 0$.  

In this particular model, one cannot write a simple expression for the balance between rate and yield.  Roughly, the idea is that a fraction $z$ of energy is put into increasing the rate of resource acquisition, and a fraction $1-z$ is put into reproduction or yield.  If the factors simply multiplied, then fitness would be $w = z(1-z)$. The change in fitness with the character $z$ gives the selective coefficient, $s$.  The change in fitness with character value is the derivative of $w$ with respect to $z$, which gives $s = 1-2z=0$, and so $\GD S = sV = 0$ implies $z^*=1/2$.  

In the actual model, the length of colony survival affects the balance between rate and yield. Short-lived colonies are favored to grow quickly (high rate and low efficiency) to use up available resources before extinction, whereas long-lived colonies are favored to grow slowly and use resources efficiently.  Thus, in the lower curve of \Fig{microbe}, longer colony survival causes the optimal balance to shift toward lower rate and higher yield.

\subsection*{Balance between selection and transmission}

When mutation generates variation within colonies, then the rate-yield tradeoff balances selection between colonies and the transmission bias from selection within colonies.  The upper two curves in \Fig{microbe} show the equilibrium balance, $z^*$.  The top curve has a mutation rate ten times greater than the middle curve.  

As colony survival increases, the equilibrium moves toward greater investment in resource acquisition. Higher resource acquisition and metabolic rate arise from the inevitable production of mutant neighbors within colonies and  the multiple rounds of internal selection within groups.  With very long colony survival times, both upper curves would converge to a high value of $z^*$ at which nearly all resources are devoted to resource acquisition and competition within colonies, with the yield efficiency dropping to a very low level.  At that point, transmission bias from selection within groups dominates selection bias between groups.

The equilibrium rate-yield tradeoff reflects the fundamental balance between selection and transmission.  That balance provides a simple conceptual basis for understanding how natural selection shapes characters.  However, in this relatively realistic model, one cannot use the balance of \Eq{totalBalance} directly to calculate the predicted outcome.  Instead, I had to use other mathematical methods to obtain the solution \autocite{frank10the-trade-off}. The selection-transmission balance only provides a framing in which to interpret the results.  

In the earlier models in this article, it was easy to calculate the ratio of selection to transmission, $\R$.  Here, the calculation is difficult, and methods such as the Price equation, kin selection, and group selection are of no use in calculating the outcome.  After obtaining a solution by other means, one can use those framings to interpret the forces that shaped the outcome.  This limitation to post hoc explanations is typical of the grand theories when faced with realistic scenarios.  Across the range of different problems presented in this article, the selection versus transmission framing provides the most general conceptual view, following the spirit of the epigraph by Williams. 

\section*{Conclusions}

This article is about the relative contributions of selective bias and transmission bias to overall evolutionary change.  For any problem, we first choose a higher level of organization, such as a group, an individual, or a cell within a multicellular aggregation.  Selective bias arises from differing success among the higher level entities.  Transmission bias arises from changes in character values between higher level entities and their descendants.  Transmission biases may occur by mutation, by random fluctuations, and by selection within the group.

The ratio of selective bias to transmission bias provides a simple measure for the relative dominance of the higher to the lower level of organization in overall evolutionary change.  When the two levels oppose each other, then the relative dominance of one level over the other often sets the level at which functional coherence and individuality emerges.

A key aspect of Maynard Smith and Szathm{\'a}ry's \citeyear{maynard-smith95the-major} \textit{The Major Transitions in Evolution} was expressed by \textcite[pp.~229--230]{maynard-smith88evolutionary}:
\begin{quote}
One can recognize in the evolution of life several revolutions in the way in which genetic information is organized. In each of these revolutions, there has been a conflict between selection at several levels. The achievement of individuality at the higher level has required that the disruptive effects of selection at the lower level be suppressed.
\end{quote}

Maynard Smith's suppression of disruptive effects at the lower level causes selective bias at the higher level to dominate. The quote and the conceptual basis of the major transitions therefore expresses Williams' notion of the ratio of selective bias to endogenous rate of change \autocite{michod97cooperation,michod03on-the-reorganization}.  

There is a large philosophical literature on the meaning of \textit{individuality} and of \textit{units of selection} in relation to \textit{levels of selection} \autocite{sober94a-critical,okasha06evolution}. One can certainly learn from studying that philosophical literature.  However, I have found it more instructive to analyze a wide range of interesting biological problems, to discover in practice what is actually needed to understand those problems, and to learn what general concepts link the different problems within a common conceptual basis \autocite[cf.][]{michod97evolution,michod06the-group}.  Philosophical induction from numerous evolutionary deductions. 

\section*{Acknowledgments}

My research is supported by National Science Foundation grant EF-0822399, National Institute of General Medical Sciences MIDAS Program grant U01-GM-76499, and a grant from the James S.~McDonnell Foundation.  

\bibliography{main}

\end{document}